\documentclass[11pt,a4paper]{article}
\usepackage[utf8]{inputenc}
\usepackage{amsmath}
\usepackage{amsfonts}
\usepackage{amssymb}
\usepackage{times}
\usepackage{authblk,fullpage}
\usepackage{array,graphicx}

\graphicspath{{./figures/}}

\begin{document}

\newcommand{\nwc}{\newcommand}
\nwc{\vs}{\vspace}
\nwc{\hs}{\hspace}
\nwc{\la}{\langle}
\nwc{\ra}{\rangle}
\nwc{\nn}{\nonumber}
\nwc{\Ra}{\Rightarrow}
\nwc{\wt}{\widetilde}
\nwc{\lw}{\linewidth}
\nwc{\ft}{\frametitle}
\nwc{\ben}{\begin{enumerate}}
\nwc{\een}{\end{enumerate}}
\nwc{\bit}{\begin{itemize}}
\nwc{\eit}{\end{itemize}}
\nwc{\dg}{\dagger}
\nwc{\mA}{\mathcal A}
\nwc{\mD}{\mathcal D}
\nwc{\mB}{\mathcal B}

\nwc{\Tr}[1]{\underset{#1}{\mbox{Tr}}~}
\nwc{\pd}[2]{\frac{\partial #1}{\partial #2}}
\nwc{\ppd}[2]{\frac{\partial^2 #1}{\partial #2^2}}
\nwc{\fd}[2]{\frac{\delta #1}{\delta #2}}
\nwc{\pr}[2]{K(i_{#1},\alpha_{#1}|i_{#2},\alpha_{#2})}
\nwc{\av}[1]{\left< #1\right>}

\nwc{\zprl}[3]{Phys. Rev. Lett. ~{\bf #1},~#2~(#3)}
\nwc{\zpre}[3]{Phys. Rev. E ~{\bf #1},~#2~(#3)}
\nwc{\zpra}[3]{Phys. Rev. A ~{\bf #1},~#2~(#3)}
\nwc{\zjsm}[3]{J. Stat. Mech. ~{\bf #1},~#2~(#3)}
\nwc{\zepjb}[3]{Eur. Phys. J. B ~{\bf #1},~#2~(#3)}
\nwc{\zrmp}[3]{Rev. Mod. Phys. ~{\bf #1},~#2~(#3)}
\nwc{\zepl}[3]{Europhys. Lett. ~{\bf #1},~#2~(#3)}
\nwc{\zjsp}[3]{J. Stat. Phys. ~{\bf #1},~#2~(#3)}
\nwc{\zptps}[3]{Prog. Theor. Phys. Suppl. ~{\bf #1},~#2~(#3)}
\nwc{\zpt}[3]{Physics Today ~{\bf #1},~#2~(#3)}
\nwc{\zap}[3]{Adv. Phys. ~{\bf #1},~#2~(#3)}
\nwc{\zjpcm}[3]{J. Phys. Condens. Matter ~{\bf #1},~#2~(#3)}
\nwc{\zjpa}[3]{J. Phys. A ~{\bf #1},~#2~(#3)}
\nwc{\zpjp}[3]{Pramana J. Phys. ~{\bf #1},~#2~(#3)}

\title{ Role of partition in work extraction from multi-particle Szilard Engine}
\author[1,2]{P. S. Pal and} 
\author[1,2]{A. M. Jayannavar}
\affil[1]{Institute of Phsyics,  Sachivalaya Marg, Bhubaneswar - 751005, India}
\affil[2]{Homi Bhabha National Institute, Training School Complex, Anushakti Nagar, Mumbai 400085, India.}
\date{} 
\maketitle
\begin{abstract}

In this work we have calculated analytically the work extraction in multi-particle Szilard engine. Unlike the previous studies, here we have introduced the biasing in the measurement procedure by inserting the
 partition at an arbitrary distance from the boundary. We found the work extraction to be symmetric with respect to a position- which is half way between the boundary walls. The work extraction is
 also calculated as a function of number of particles and it shows to saturate to a certain value for large number of  particles. We find that work extraction can be made larger for multi-particle engine
 when the partition is inserted in the middle.
 \bigskip
 
 
 \end{abstract}

\section{Introduction}
Almost 150 years ago, Maxwell put forward a thought experiment thereby threatening the validation of second law of thermodynamics. He introduced a demon like creature which controls the dynamics of 
an isolated system and while doing so the entropy of the  system decreases\cite{max,vedral_09}. This violates the second law. A classical analysis of Maxwell's demon was conducted by Szilard \cite{Szilard}
where he studied an 
idealized heat engine with one particle gas and directly associated  the information acquired by measurement with the physical entropy and saved the second law. Introduction of Maxwell's demon has 
stirred the physics community working in information themodynamics  and has lead to some insightful physics \cite{landauer_61,landauer_91,Esposito_11,mandal_12,mandal_13,parrondo_15,rana_16,rana_1611}. Quantum Szilard engine(QSZE) was first studied
by Kim et al. \cite{kim_11} where they have clearly analyzed the quantum effects. In their paper, they have shown that- unlike classical case- insertion and removal of partition in QSZE leads to some amount of work
done on the system. In fact, the work extraction also depends on the type of particles - Bosons or Fermions- used as working substance. Many particle QSZE had been extensively studied in \cite{Lu_12},
 where the total work is shown to be dependent on particle statistics, the odd-even parity and the temperature of the system. The parity effect is seen to affect the system when the working substance 
 is a collection of Fermions. An odd number of Fermions performs work whereas an even number of Fermions do not perform work at all. For Bosonic QSZE, there exists a critical temperature below which
 the engine cannot perform work i.e., the work is negative. As shown in this paper, that above the critical temperature the Bosonic QSZE performs more work than Fermionic QSZE. In \cite{kim_16}, 
 optimization of work extraction in QSZE has been discussed for multi-particle case. They have shown that multi-particle QSZE has an inevitable irreversibility which makes the optimization challenging.
 In this brief note, we are studying the work extraction in  classical multi-particle Szilard engine with a biasing. To our knowledge Szilard engine is studied with the partition being initially
 put in the middle of the box i.e., the partition divides the phase space of the working substance into two equal halves. In our case this division of phase space is made arbitrary by inserting the 
 partition at an arbitrary distance from the boundary. This action changes the probability of particles to stay in certain volume. In this  sense we are calling it a biasing. Our aim is to see how 
 this affect the work extraction. Another important question is how this biasing can lead to an optimal work extraction in multi particle Szilard engine. We find that work extraction can be made
 larger for M-particle engine when the partition is inserted in the middle.

 In the next section we will describe the many particle classical Szilard Engine in detail. Section 3 describe the work extraction for multiparticle classical Szilard 
 engine and how it depends on the biasing and the number of particles. Finally, we will conclude. Each section is made self consistent.

 \section{Szilard Engine}
 We consider a Szilard engine(SZE) consisting of $M$ particles in a closed cylinder of horizontal length $L$. There are four stages in the thermodynamic cycle of a SZE namely - (i) inserting a wall
 in the vertical direction at $l=xL(0\leq x \leq 1)$, (ii) measuring the number of particles $n$ on the left side, (iii) performing an isothermal expansion in contact with a heat bath at temperature
 $T$, where the wall stops at $l=l_n$ differing in $n$, and (iv) removing the wall to complete the cycle. The work done by the SZE is given by[]
 
 \begin{equation}
  W_{tot}=-k_BT\sum_{n=0}^M p_n\ln\left(\frac{p_n}{f_n}\right).
  \label{w_tot}
 \end{equation}
$k_B$ denotes the Boltzmann constant which is taken to be unity in subsequent expressions. Here $p_n$ and $f_n$ are given by
 \begin{equation}
 p_n=\frac{Z_{n,M-n}(xL)}{\sum_{n'}Z_{n',M-n'}(xL)};f_n=\frac{Z_{n,M-n}(l_n)}{\sum_{n'}Z_{n',M-n'}(l_n)},
\end{equation}
 where $Z_{n,M-n}(X)=Z_n(X)Z_{M-n}(L-X)$ is a partition function that describes the situation of $n$ particles to the left of the partition and the remaining $M-n$ to  the right, in a 
thermal equilibrium. $Z_n(X)=cX^n/n!$ is the partition function of $n$-particle in a box of length $X$ and $c$ is some constant. Physically, $p_n$ denotes the probability that there are $n$ particles to the left after partition and $f_n$ represents the probability to choose the case of $n$ particles on the 
left side of the wall when the wall is inserted at $l_n$ in the time backward process. Note that one can choose $l$ freely when the wall is inserted while $l_n$ is determined from the force 
balance on both sides of the wall $F^L+F^R=0$. 

Although Eq. \ref{w_tot} is derived for QSZE, surprisingly, it also holds true for work done by classical SZE derived using classical non-equilibrium thermodynamics. But for both the cases the 
partition function differ and so does the amount of work.

\section{Work extraction in classical Szilard engine}

In classical Szilard engine, no work is done on the system during insertion and removal of partition. Quasistatic and isothermal shifting of partition contributes to the work in the whole cycle. 
%
%
%
In case of system consisting of $M$ particles, $p_n={M \choose n} x^n(1-x)^{M-n}$ and $l_n=(n/M)L=rL$. 

\begin{eqnarray}
f_n&=&\frac{Z_{n,M-n}(rL)}{\sum_{n'}Z_{n',M-n'}(rL)}, \nn\\
&=&\frac{Z_n(rL)Z_{M-n}((1-r)L)}{\sum_{n'}Z_{n'}(rL)Z_{M-n'}((1-r)L)},\nn\\
&=&\frac{\frac{(rl)^n}{n!}\cdot\frac{[(1-r)L]^{M-n}}{(M-n)!}}{\sum_{n'}\frac{(rl)^{n'}}{n'!}\cdot\frac{[(1-r)L]^{M-n'}}{(M-n')!}},\nn\\
&=&{M \choose n}\frac{(rL)^n[(1-r)L]^{M-n}}{L^M}={M \choose n}r^n(1-r)^{M-n}.
\end{eqnarray}

\begin{figure}
 \begin{center}
  \includegraphics[height=2.5 in,width=2.5 in, angle=270]{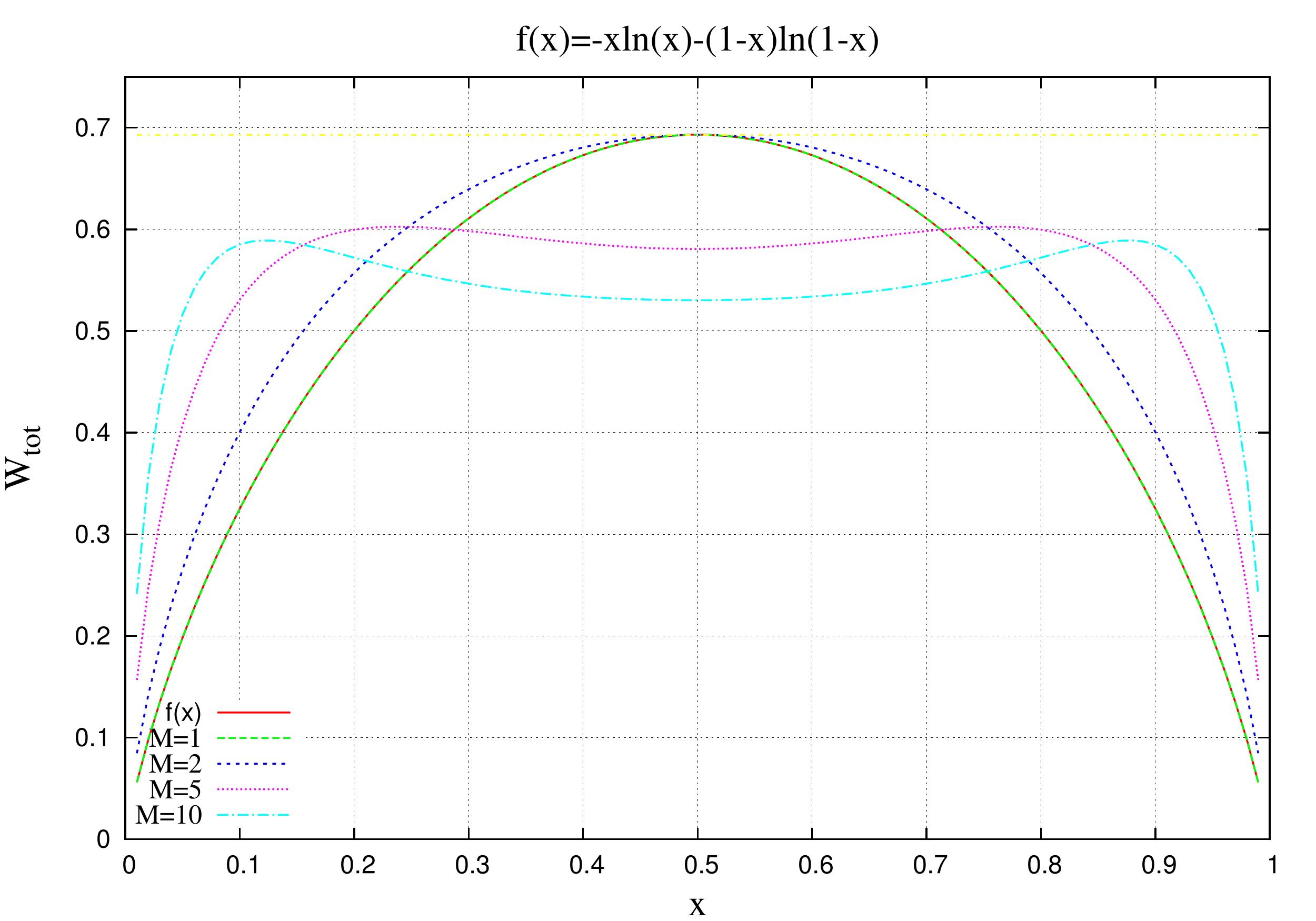}
  \caption{ Plot of work done on the system as a function of the initial position of the partition.}
  \label{W_vs_x}
 \end{center}

\end{figure}

Inserting  these expressions of $p_n$ and $f_n$ in Eq.\ref{w_tot}, we obtain 

\begin{equation}
 W_{tot}(x,M)=-T\sum_{n=0}^M {M \choose n} x^n(1-x)^{M-n}\ln\left[\left(\frac{x}{r}\right)^n \left(\frac{1-x}{1-r}\right)^{M-n}\right].
 \label{w_tot1}
\end{equation}
Fig.\ref{W_vs_x} depicts the behavior of work done by SZE as a function of biasing for different number of particles. It is seen right away that the work extraction is symmetric about $x=0.5$, which  
is quite obvious. Another interesting fact to note is the work done by SZE is same for $M=1$ and $M=2$ when the wall is initially kept at $x=0.5$; it is different otherwise. This is due to the fact that
when there is no biasing, the measurement outcome that there is one particles on both sides does not contribute to the work extraction as the system stays at equilibrium with the wall at $x=0.5$ 
and it does not move. This is not case for $x\neq 0.5$. Work extraction has a single peak at $x=0.5$ for single particle but it splits into two distinct peaks as the number of particles increases. 
This gives clear clue that work extraction is more for multiparticle working substance with biasing.

\begin{figure}
 \begin{center}
  \includegraphics[height=2.5 in,width=2.5 in, angle=270]{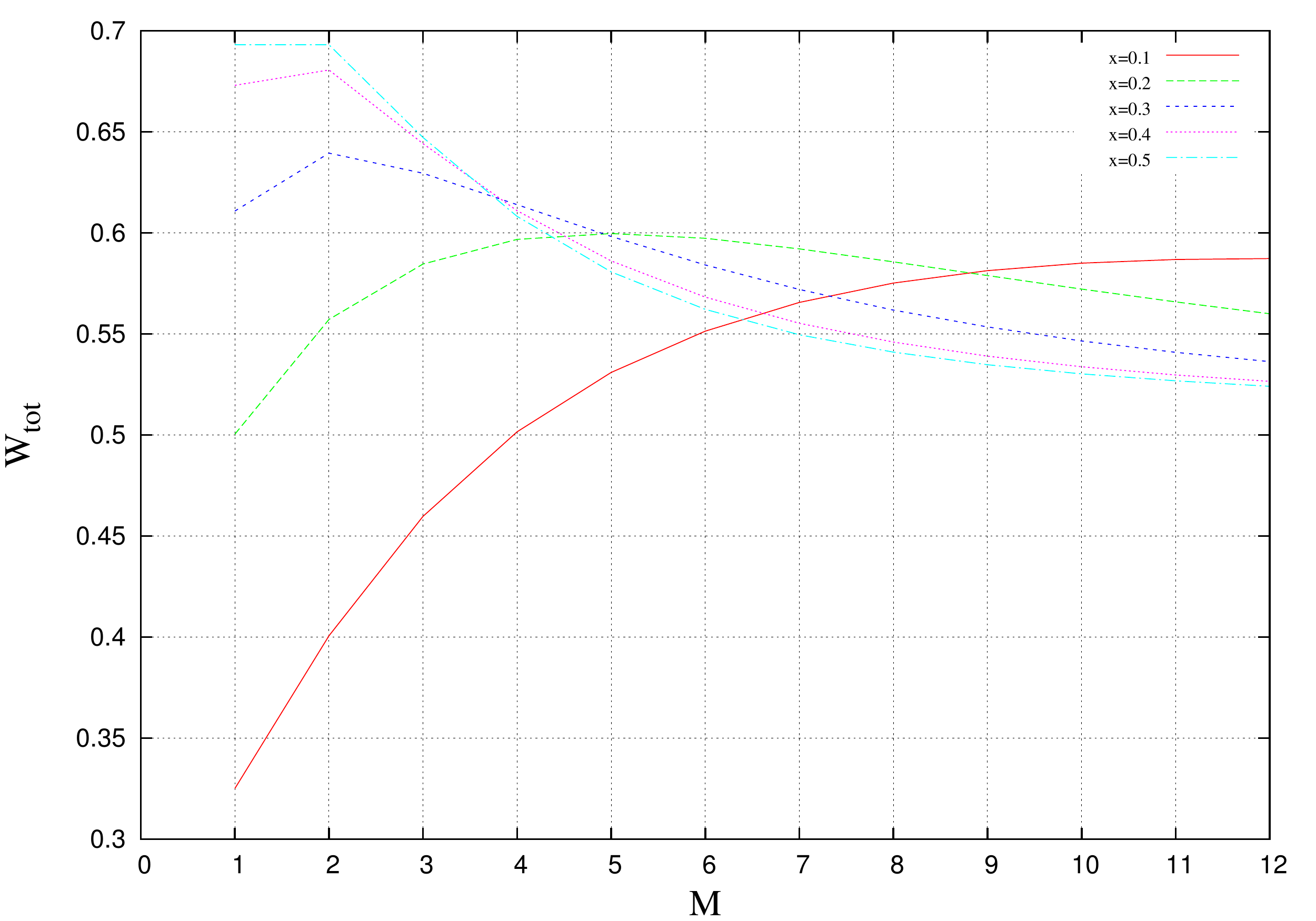}
  \caption{ Plot of work done on the system as a function of number of particles in the system.}
  \label{W_vs_M}
 \end{center}

\end{figure}

Fig.\ref{W_vs_M} shows the work extraction as a function of number of particles for different biasing. The plot depicts the fact that work extraction is small with large biasing when number of
particles is small. But work done by SZE increases for large biasing when the particle number is large.

\section{Conclusion}
In conclusion, we have extensively studied multi-particle classical Szilard engine with biasing. We have shown that work extraction can be enhanced for large number of particles by inserting the partition
at a different position other than the mid-point. Presently we are working on quantum Szilard engine. 

\section{Acknowledgement}
We thank Prof. T. Sagawa for clarifying some aspects of quantum Szilard engine. One of us (AMJ) thanks DST, India for financial support (through J. C. Bose National Fellow-ship).

\end{document}